\documentclass[a4paper,fleqn,usenatbib]{mnras}
\usepackage[T1]{fontenc}
\usepackage{ae,aecompl}
\usepackage{graphicx}	
\usepackage{amsmath}	
\usepackage{amssymb}	

\title{Cosmic-Ray Interactions in the Solar Atmosphere} 

\author[Hudson et al.]{
\begin{tabular}{lll}
Hugh S. Hudson$^{1,2}$,
Alec MacKinnon$^{1},$
Mikolaj Szydlarski$^{3}$,
and Mats Carlsson$^{3}$
\end{tabular}
\bigskip
\\
$^{1}$School of Physics and Astronomy, U. of Glasgow, UK\\
$^{2}$Space Sciences Laboratory, UC Berkeley, CA USA\\
$^{3}$Institute of Theoretical Astrophysics, University of Oslo, Norway
}

\begin{document}
\label{firstpage}
\pagerange{\pageref{firstpage}--\pageref{lastpage}}
\maketitle

\begin{abstract}
High-energy particles enter the solar atmosphere from Galactic or solar coronal sources, producing an ``albedo'' source from the quiet Sun, now observable across a wide range of photon energies.
The interaction of high-energy particles in a stellar atmosphere depends essentially upon the joint variation of the magnetic field and the gas, which heretofore has been characterized parametrically as $P \propto B^\alpha$ with $P$ the gas pressure and $B$ the magnitude of the magnetic field.
We re-examine that parametrization by using a self-consistent 3D MHD model (Bifrost) and show that this relationship tends to $P \propto B^{2.9\pm0.1}$ based on the visible portions of the sample of open-field flux tubes in such a model, but with large variations from point to point.
 This scatter corresponds to the strong meandering of the open-field flux tubes in the lower atmosphere, which will have a strong effect on the prediction of the emission anisotropy (limb brightening).
The simulations show that much of the open flux in coronal holes originates in weak-field regions within the granular pattern of the convective motions seen in the simulations.
\end{abstract}


\section{Introduction}

Galactic cosmic rays permeate the heliosphere, and can penetrate to the surface of the Sun.
The unique way of remotely sensing this process so far has been the detection of $\gamma$-radiation from secondary interactions.
In the meanwhile direct explorations of the inner heliosphere via the \textit{Parker Solar Probe} and the forthcoming \textit{Solar Orbiter} will soon become available.
The recent \textit{Fermi} observations of high-energy ($> 20$~MeV) $\gamma$-rays have revealed not one, but at least two distinct solar $\gamma$-ray sources resulting from the interactions of Galactic cosmic rays (GCRs): one from the disk of the Sun, and one from its extended corona \citep{2011ApJ...734..116A}.
The latter results from Compton scattering of GCR electrons on the solar photon field, but explaining the disk component requires much more complicated theoretical work, tracking the GCR ions through the heliosphere and down into the solar atmosphere along the magnetic field.

At the surface of the Sun, the solar high-energy community typically has used rudimentary models of the atmosphere to study cosmic-ray interactions.
Nowadays we have self-consistent MHD models that describe the complex structure of the magnetized solar atmosphere.
These replace the traditional 1-D semi-empirical models such as VAL-C \citep{1981ApJS...45..635V}, both geometrically and also because the MHD models actually contain a magnetic field.
This presents a very different scenario for the particle interactions, with consequences for the $\gamma$-ray luminosity of the Sun and for its image structure in terms of spatial distribution and directivity (limb brightening, observationally).

At the Earth the equivalent analysis involves studying secondary effects from the interactions GCRs in the terrestrial atmosphere, detected in several ways but notably via the network of neutron monitors \citep{2000SSRv...93...11S}.
The pioneering work of \cite{Stoermer...1955} described the propagation of cosmic rays in the Earth's magnetic environment, but so far as we are aware there has been no comparable development for the global magnetic field of the Sun and inner heliosphere.
Despite this, observations of TeV $\gamma$-rays show clear shadows due to the presence of the Moon and the Sun, and the solar shadow clearly reflects the presence of the magnetic field in these regions \citep{1993ApJ...415L.147A}.
These considerations suggest that the ``solar St{\o}rmer problem'' will have extremely interesting developments that bear upon problems of major importance in both solar and cosmic-ray physics.
The GCR particles at the Sun have already revealed themselves via \textit{Fermi} observations \citep{2014ApJ...787...15A,2018ApJ...869..182S}, following the initial high-energy $\gamma$-ray observations of the quiet Sun \citep{2008A&A...480..847O}.

In addition to the galactic cosmic rays (GCRs), of course, the solar energetic particles (SEPs) interact with the solar magnetic field in more complicated ways, indeed requiring sudden restructurings of the coronal magnetic field to come into existence at all.
As with the quiet-Sun $\gamma$-rays due to GCRs \citep{2008A&A...480..847O}, earlier observations \cite[e.g.,][]{1995ARA&A..33..239H} laid the groundwork for the SEPs interactions at the Sun \citep{1991ApJ...382..652S}.
We remark that \cite{2014ApJ...787L..25R} have analyzed the GCR interactions on Jupiter, noting that either $\gamma$-radiation or gyrosynchrotron radiation from extensive air showers may be detectable.
The Jovian environment is more Sun-like than Earth-like from the point of view of elemental abundances, but unlike the Sun it appears to have a smooth large-scale magnetic field.

In this paper we deal with a restricted part of the broad issues involved with GCR and SEP interactions at the Sun.
Specifically we consider the details of the environment within which lower atmosphere particle interactions take place, based upon newly available models 
that self-consistently describe the MHD activity in this region.
Such models differ radically from the 1D semi-empirical atmospheric models \citep[e.g.][]{1981ApJS...45..635V} long in use for a variety of problems. These do not even include a magnetic field so they can only provide a background density and temperature structure on which a magnetic field model, derived separately from quite separate considerations, can be imposed. 
The commonly used approximation for the magnetic structure has been the parametrization proposed by \cite{1983ApJ...264..648Z}, namely that the gas pressure $P \propto B^\alpha$, with the exponent $\alpha$ treated as a free parameter.
We ``calibrate'' this relationship with Bifrost models \citep[e.g.][]{2016A&A...585A...4C} but restrict our attention, for simplicity, to a model describing the atmospheric structure in a coronal hole. 

Coronal holes have a particular interest both for GCRs and for SEPs.
In the accepted pictures of these particles, the GCRs arrive at the Sun via the ``open'' heliospheric magnetic fields, thought to concentrate in the coronal holes.
The SEPs have a strong association with large-scale shock waves driven by coronal mass ejections (CMEs); these can accelerate particles within open-field regions, and so at least speculatively the coronal holes should become bright during an SEP event (Hudson 1988, cited in Seckel et al., 1991).
\nocite{1991ApJ...382..652S}
This speculation has not been confirmed yet, and the limited information on source localization provided by \textit{Fermi} hints that the picture may not be so simple \citep[e.g.][]{2018ApJ...865L...7O}.

\section{Model field behavior}

\subsection{A Bifrost model of open-field footpoints}

\begin{figure*}
\includegraphics[width=2\columnwidth]{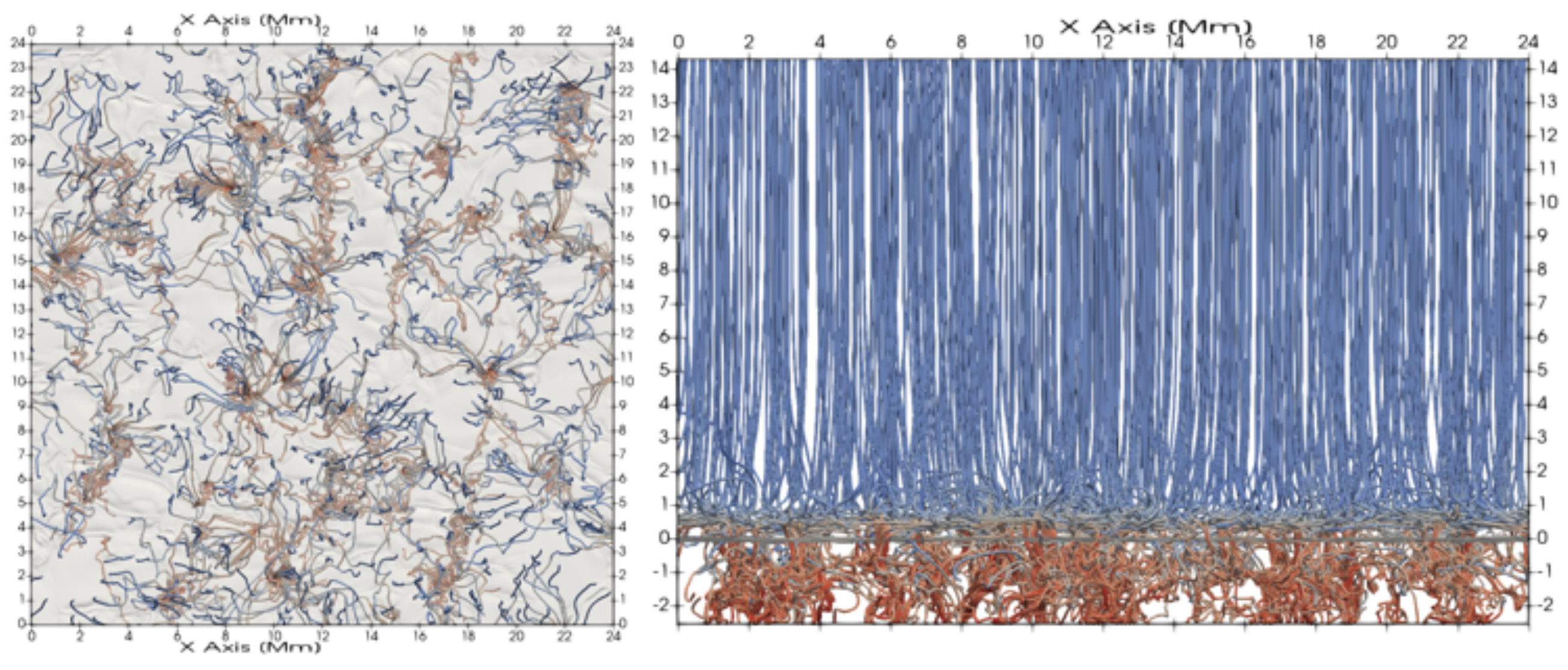}
\caption{Two views of the ``open'' fluxtubes in the Bifrost model Bifrost model: left, from above, and right, in the Y direction.
Note the thinness of the interface region above Z = 0, where most of the radiation output forms.
     }
\label{fig:bifrost_open}
\end{figure*}

We have taken a particular standard Bifrost model (``BIFROST\_en024048\_hion'') to represent the quiet Sun in a coronal hole.
This model ensures an open flux by requiring an unbalanced mean field of 5~G; the computational box has 24~Mm resolution horizontally and about 48~Mm vertically, but with a non-uniform vertical grid.
Figure~\ref{fig:bifrost_open} illustrates the  field structure at one time step after the model has relaxed to a steady state. 

The Bifrost models have a rectilinear configuration and nominally extend from inside the convection zone up into the solar corona, in this case including heights in the range [-2.19630, 14.2887]~Mm relative to the photosphere, defined as $\tau \approx 1$ at 5000~\AA\ and covering a horizontal area  of 24$\times$24~Mm at a grid spacing of 48~km.
Without a more complete simulation one cannot know exactly what field would correspond to the solar wind, but the adopted signed flux imbalance corresponds to the requirement for the total open flux known observationally at one A.U. \citep[e.g.][]{2013LRSP...10....4L}.

The model successfully reproduces a granulation pattern, but does not have a large enough scale for the supergranulation and the chromospheric network.

For the discussion below, we have taken 512 random points near the upper surface of the datacube, at Z~=~14.0~Mm, and followed them to  Z~=~0~Mm.
These constitute the open fieldlines that would guide cosmic rays incident from larger scales in the heliosphere.
A further restriction to eliminate lines crossing the side walls of the datacube resulted in a selection of 441 total.

\subsection{Cosmic-ray transport near the Sun}

We know almost nothing about cosmic-ray transport near the Sun.
In the heliosphere at large, cosmic-ray transport appears to follow a diffusion theory, which can successfully explain solar-cycle modulation and Forbush decreases, for example \citep[e.g.][]{2013LRSP...10....3P}.
One could in principle extrapolate the parameters of the existing transport theory to the vicinity of the Sun, the approach taken by \cite{1991ApJ...382..652S,1992NASCP3137..542S}.
At the Earth itself, the St{\o}rmer calculation of cosmic-ray cutoff rigidities consists of test-particle integrations of particle dynamics from a distant source, penetrating a specified field geometry.

In contrast to a purely diffusive theory for cosmic-ray transport, one has clear signatures of deterministic transport, for example in the prompt arrival of flare-associated SEPs and the presence of non-dispersive boundaries in particle events \citep[e.g.][]{2000ApJ...532L..79M}.
The usual assumption at the Sun itself, for the literature describing interactions producing $\gamma$-rays and neutron secondaries,
is of simple vertical precipitation into a 1D model atmosphere such as VAL-C, supplemented by some -- usually simple -- assumptions about magnetic field strength and geometry.

\begin{figure}
\includegraphics[width=\columnwidth]{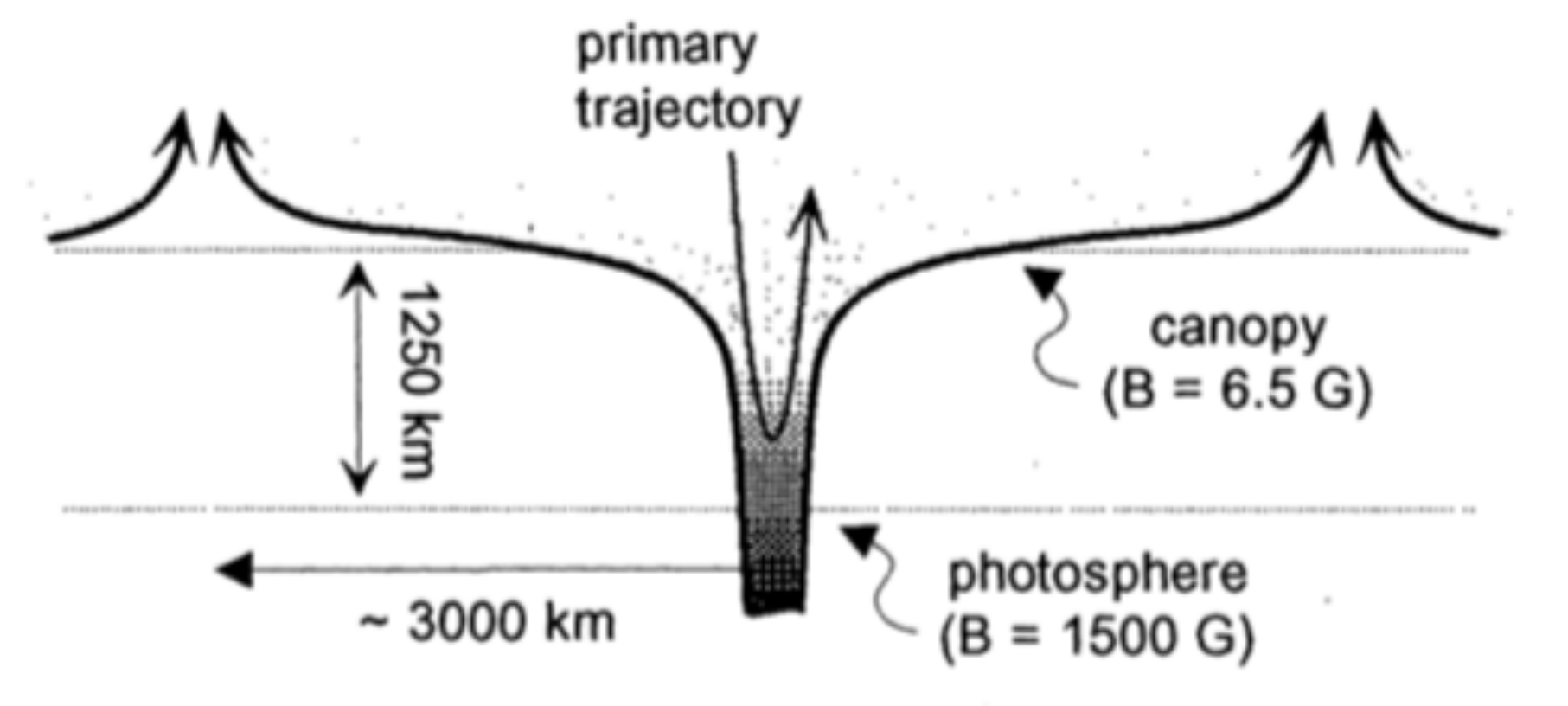}
\caption{The canopy geometry assumed by Seckel \textit{et al.} (1992).
     }
\label{fig:seckel}
\end{figure}

The substantial previous work for predicting the quiet Sun's $\gamma$-ray spectrum \citep{1991ApJ...382..652S} incorporated the magnetic field using a canopy geometry, as illustrated in Figure~\ref{fig:seckel}.
\nocite{1991ApJ...382..652S}
The flux of primary particles into the dense atmosphere, in this picture, is limited by the mirror force, and the Seckel \textit{et al.} papers 
treat this as a parameter termed the absorption coefficient $A$, which they estimate to be 0.0052 for proton primaries.
This corresponds to a specific loss cone, which depends upon how dense the atmosphere is at the mirror point.
Implicit in this kind of model, but not discussed in these important papers, is the particle's direction of motion at the mirror point: a simple
vertical field would result in basically horizontal motions at the mirror point, where the maximum slowing-down (via Coulomb interactions)
and $\gamma$-radiation via nuclear interactions would occur.
These would initiate showers of secondaries as described for Jupiter by \cite{2014ApJ...787L..25R}.
Note that in the Seckel \textit{et al.} work, the authors assume $P \propto B^2$, corresponding to simple pressure balance and to $\alpha = 2$ in the Zweibel-Haber parametrization.

\subsection{Larmor motion}

Charged particles gyrate around the magnetic field, but must follow it closely if the field gradients are small relative to the Larmor radius $R_L$ of the gyration, which depends upon particle energy and $B$.
The Bifrost numerical cell size, at which level the simulation must be smooth, corresponds to the gyroradius of a 6~GeV proton at 5~G, and so the assumption of adiabatic particle motion will fail somewhere above the heart of the GCR spectrum (of order 1~GeV, due to solar modulation). 
For the purposes of this paper, however, we will continue with the idea that the particles stick to the field and have negligible cross-field motion.
Results described below will help to justify this assumption.
\nocite{1992NASCP3137..542S}

\section{Implications for $\gamma$-ray observations}

\subsection{Particle interactions}

\begin{figure}
\includegraphics[width=\columnwidth]{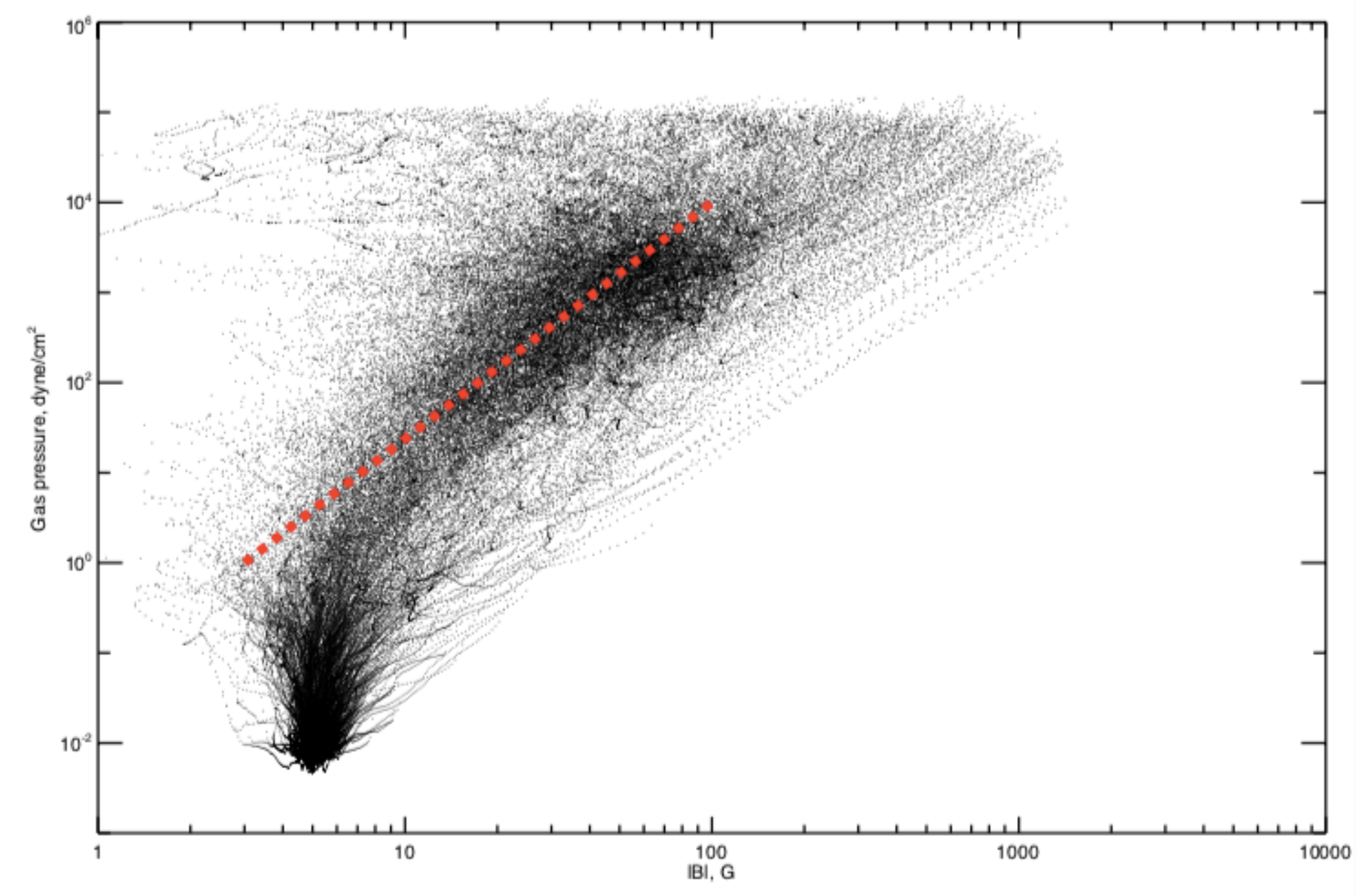}
\caption{Traces of gas pressure and total magnetic field for the open fields, as defined in the text.
The dense photosphere is at the top of the diagram, and the
red dashed line shows $\alpha = 2.67$ in the Zweibel-Haber parametrization $P \propto B^\alpha$.
     }
\label{fig:p_vs_b}
\end{figure}

Near the top boundary of the simulation domain, at $Z = 14$~Mm), we select a random set of points.
As seen in Figure~\ref{fig:bifrost_open}, this corresponds to a  reasonable definition of open field.
We follow the field from each of these points downward, calculating the column density to compare with the particle range-energy relationship, and record the total magnetic field intensity $B$ along the path along with the gas pressure.
This lets us compare with the Zweibel-Haber parametrization.
The results for this open-field tracing appear in Figure~\ref{fig:p_vs_b} here.

The behavior of the (P, B) relationship for individual flux tubes deviates strongly from the simple Zweibel-Haber parametrization, due to the incessant motions throughout the chromosphere and photosphere. 
In this region the cosmic-ray particles arriving at the Sun slow down and stop, sometimes undergoing a nuclear reaction producing high-energy secondary products, including the observable $\gamma$ rays.
In Figure~\ref{fig:p_vs_b} one can see several examples of flux tubes that exhibit a simple power-law behavior, but many others that show great complexity, defying any simple description.
A model-based estimation of a representative value for the power-law index may not work well for some important applications, such as the estimation of the $\gamma$-ray limb-darkening law.
We return to this in Section~\ref{sec:pred}.

\begin{figure}
\centering
\includegraphics[width=0.8\columnwidth]{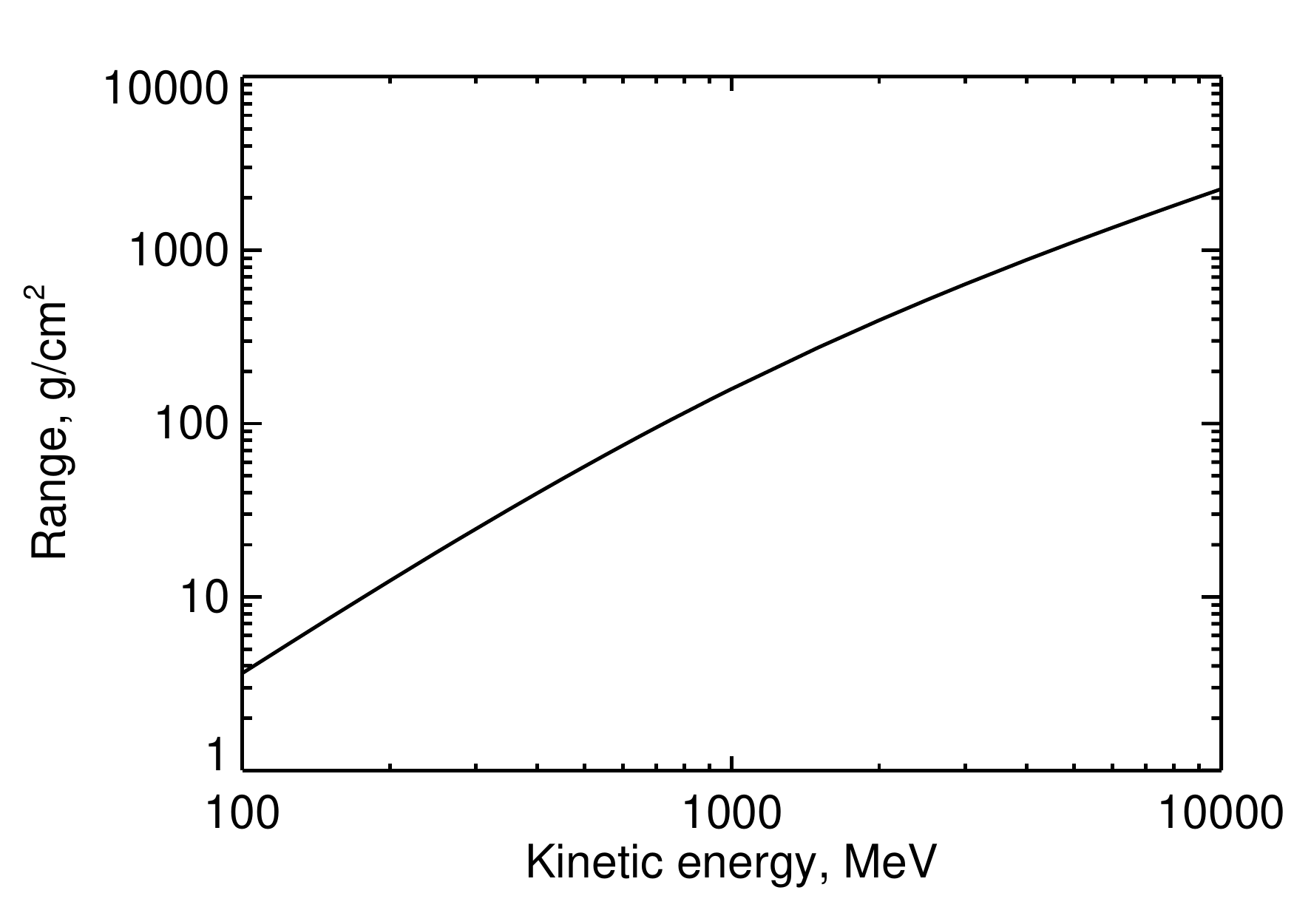}
\caption{Standard (NIST/PSTAR) range-energy relationship for protons in hydrogen.
     }
\label{fig:range-energy}
\end{figure}

We can explore this behavior more explicitly by integrating column depth (``grammage'' or surface density) along the field, since this determines the collisional losses of a cosmic-ray particle; and also the path along which it can produce nuclear interactions.
As noted in the early literature \cite[e.g.][]{1970SoPh...13..471S}, protons even at low cosmic-ray energies of a few hundred MeV can penetrate to and below the photosphere in a semi-empirical atmospheric model.
In the Earth's atmosphere the mean nuclear interaction length is about 60~g~cm$^{-2}$, corresponding to the ``Pfotzer Maximum'' of hard radiations; for the case of the Sun the interaction physics is different, but we would expect the primary interaction point to lie well below the photosphere even on rectilinear vertical precipitation.
For reference, Figure~\ref{fig:range-energy} shows the range of 0.1-10~GeV protons in hydrogen.

For predicting the mean image properties of the $\gamma$-ray albedo secondary to the cosmic rays, in the quiet Sun, one could carry out exact calculations of particle motion and interaction in an instantaneous model snapshot.

\subsection{The Zweibel-Haber parametrization}

\begin{figure}
\includegraphics[width=\columnwidth]{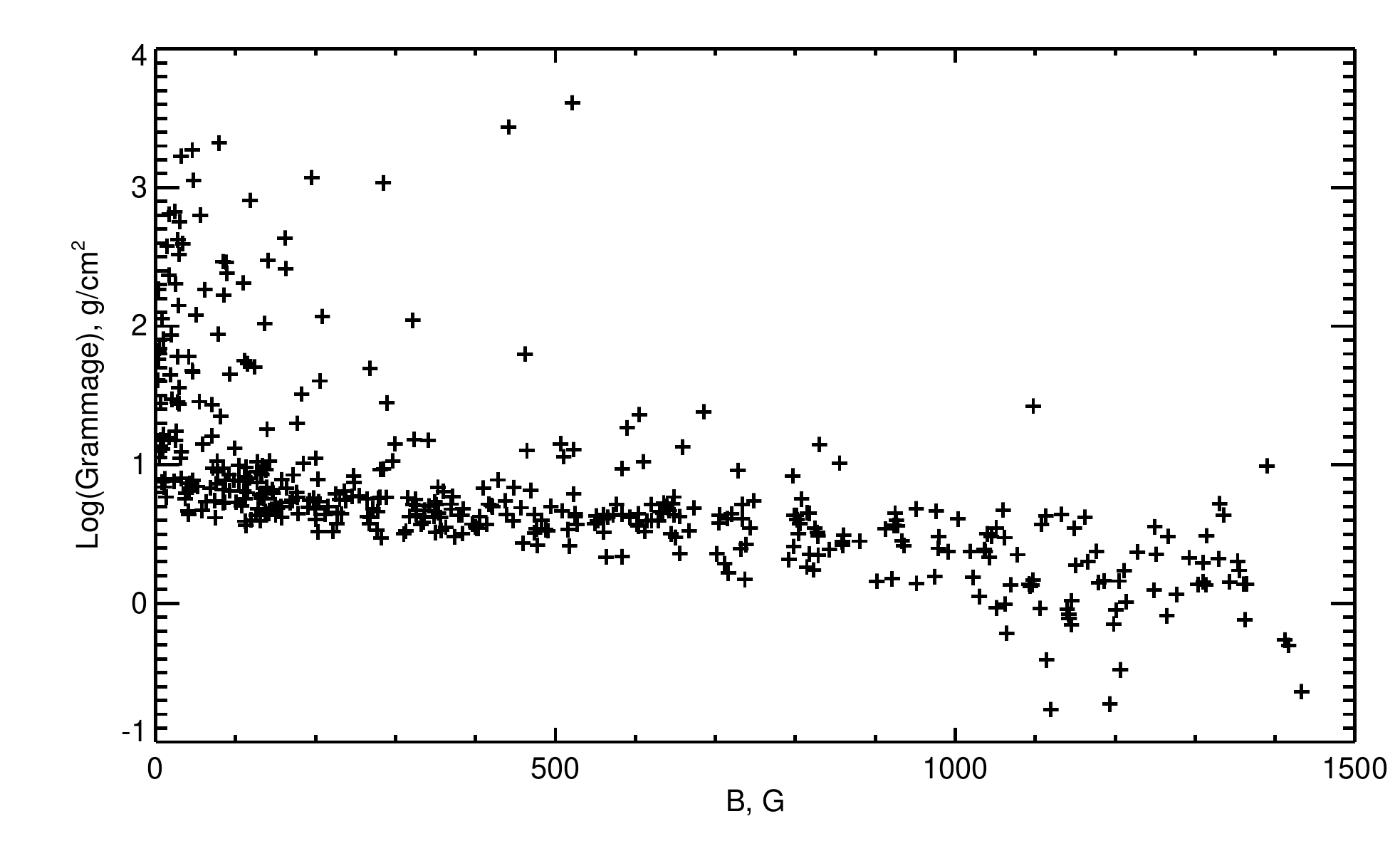}
\caption{Column density above Z = 0 on the open field lines, as a function of $B$ at Z=0 in the model.
     }
\label{fig:gram_vs_b}
\end{figure}

The Bifrost results for grammage are given here in Figure~\ref{fig:gram_vs_b}, shown as a function of footpoint field strength $B$.
They reveal a weak anti-correlation between photospheric field strength and column density, with interesting outliers.
In a sense, we expect this by analogy with the Wilson effect, interpreted as a true evacuation of flux tubes with high magnetic pressure.
The extremely low values of grammage at high $B$ seem surprising because they imply cosmic-ray access deep below the photosphere, but with the penalty of reduced parameter space because particles incident from above with large pitch angles will mirror.
Equally interesting, the large column depths seen in many low-$B$ field lines imply that a much greater population of high-energy particles can interact at chromospheric temperatures and densities than would be expected from a 1D model, independent of the mirroring restriction.
This interesting distribution could not have been seen in a 1D model without self-consistent MHD development in 3D.

\begin{figure*}
\includegraphics[width=\columnwidth]{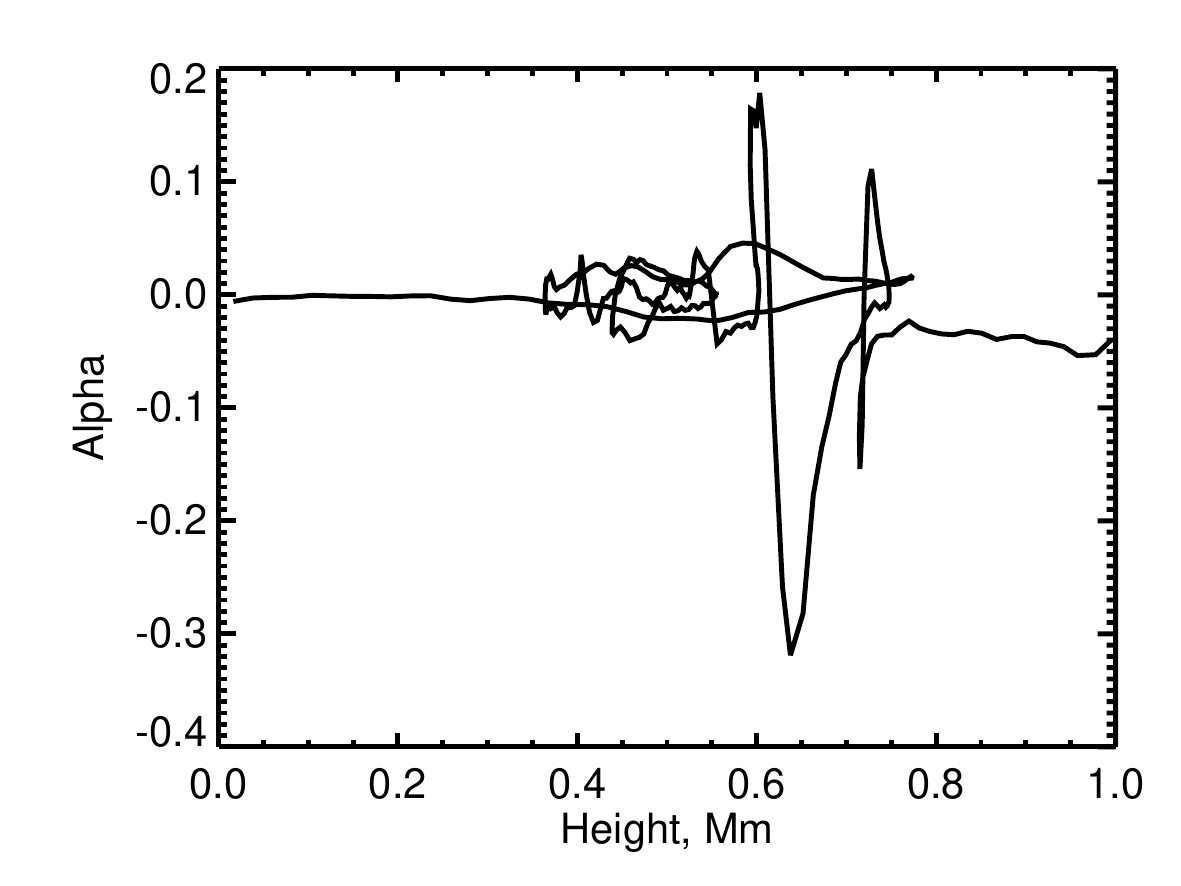}
\includegraphics[width=\columnwidth]{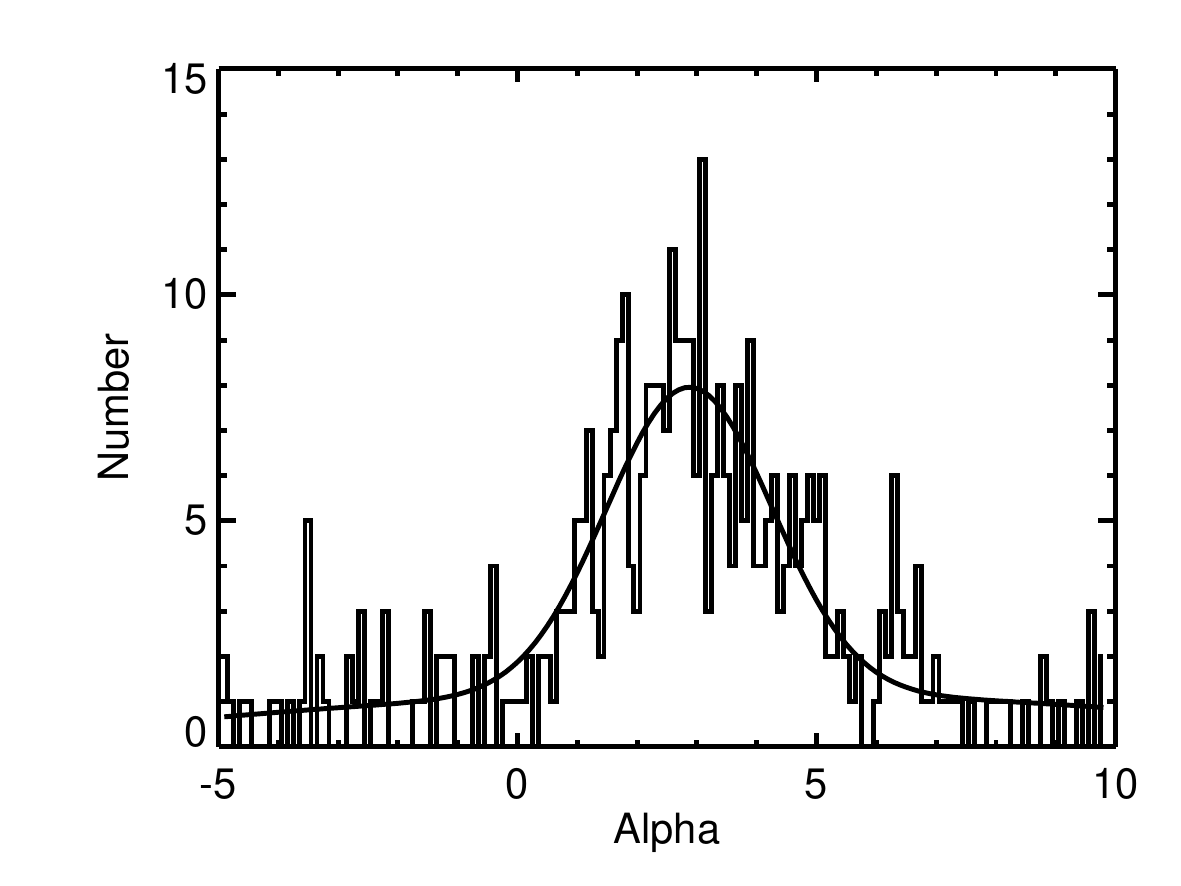}
\caption{Left, the slope $\alpha$ for $P \propto B^\alpha$ along the field direction, evaluated for one particular case.
Right, a histogram of $\alpha$ at physical height 1~Mm for the open-field set, showing a Gaussian fit for the 388/512 cases within this range.
Here we find a peak at $\alpha = 2.9 \pm 0.1$, where the uncertainty, taken from the fit, is slightly larger than the sample standard deviation.
     }
\label{fig:alpha}
\end{figure*}

Our goal here, to ``calibrate'' the Zweibel-Haber parametrization, requires the derivation of a value of $\alpha$ for the model $P \propto B^\alpha$.
Figure~\ref{fig:p_vs_b} shows a complicated relationship between these physical parameters, quite unlike what one would find in a 1D static model. 
This, no doubt, results from the complicated geometry revealed in the Bifrost models, plus their dynamism.
On a given flux tube, for a given cosmic-ray energy and initial pitch angle, there will be a distributed production of nuclear products at different points along the path, which may not even be monotonic in gas pressure.
The left-hand panel of Figure~\ref{fig:alpha} shows the variation of $\alpha$ along the field for a specific case, for the points in the height range [0, 1]~Mm.
Its great complexity results from the nearly horizontal meanderings of the field in this height range, as also seen in the upper region of Figure~\ref{fig:p_vs_b}.

The right-hand panel of Figure~\ref{fig:alpha} shows the $\alpha$ value evaluated at a height of 1~Mm for each of the 512 sample fieldlines.
Only 388 of these resulted in $-5 < \alpha < 10$, but for these a peak appears at $\alpha \approx 2.9 \pm 0.1$.
The uncertainty here is from the Gaussian fit; we think that there is so much scatter that this parametrization
has little physical significance.
The fact that the peak lies not far from the equipartition value $\alpha = 2$ probably reflects the presence of some
parts of the model domain with relatively stable properties.

\subsection{The solar wind}

\begin{figure}
\includegraphics[width=\columnwidth]{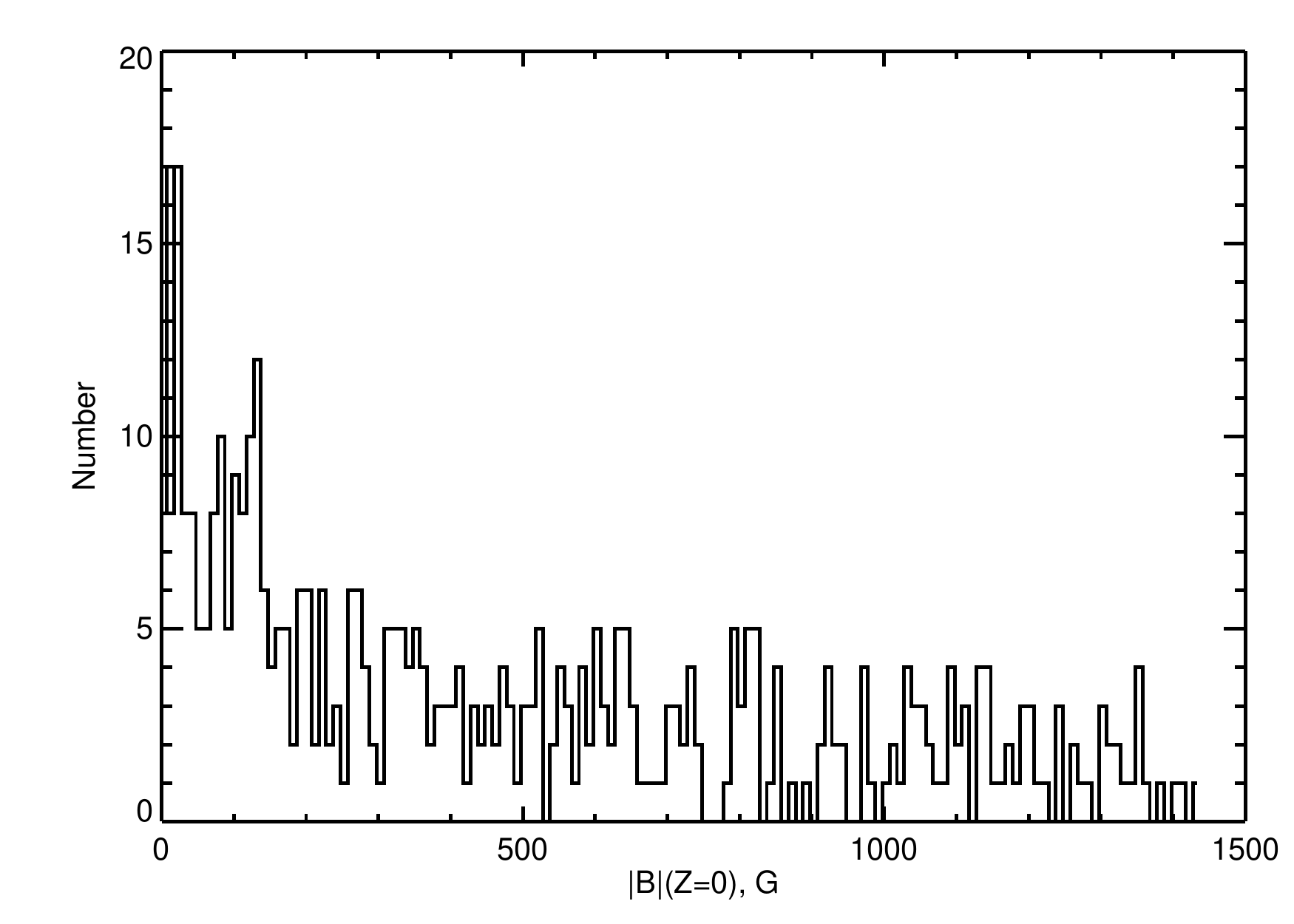}
\caption{Histogram of field magnitude at the footpoints of open fields in the model studied.
     }
\label{fig:histogram_deep}
\end{figure}

Figure~\ref{fig:histogram_deep} shows a histogram of footpoint magnetic field values for the open field set, with the somewhat surprising result that much of the open field appears to come from the weak-field regions in the model.
This would seem to contradict the general picture of the canopy behavior expected of the field (Figure~\ref{fig:seckel}).
As the granulation flow fields intensify the photospheric field into the vertices, we would expect that the strongest fields would reconnect and indeed form a canopy.
The equilibrium connectivity of the model field thus has a dependence upon the microphysics of the reconnection process, which the model can only deal with approximately.
Another possibility is that small errors in the mapping in the domain [0, 14]~Mm have allowed some artificial crosstalk between field lines.
We do not think it impossible that some of the open magnetic field emerging from a coronal hole actually \textit{does} originate in the weak
fields interior to the cell boundaries and suggest this as an area for future research.

Whatever the reason, this finding would imply much smaller mirror ratios than found by \cite{1992NASCP3137..542S} and thus a much greater fraction of the incident cosmic rays interacting to produce secondaries. More quantitative study is needed to determine if this can be part of the explanation for the high-than-anticipated quiet Sun $\gamma$-ray fluxes found by \cite{2011ApJ...734..116A}. 

\subsection{Predictions for $\gamma$-ray imaging}\label{sec:pred}

The complex meandering motions of the open-field flux tubes, as they enter the dense lower atmosphere, means that knowledge of the initial arrival direction will have largely disappeared. 
The pitch-angle distributions of the particles significantly affect the intensity of the secondary $\gamma$-ray and neutron emissions \citep[e.g.][]{1975SSRv...18..341R}, and so some of this work will need to be revisited.
In a mainly vertical magnetic field, a mirrored particle will move mainly horizontally, leading to horizontal shower development and strong limb brightening, as predicted \citep{1973NPhS..242...59S} and observed \citep{2009PhRvD..80l2004A} in the Earth's atmosphere.
In the solar atmosphere, the chaotic variations of the field direction in the mirrror points will reduce this effect.
It is beyond the scope of this paper to deal with this complicated question, but Bifrost or similar models do make it possible now to assess the distribution numerically.

\section{Conclusions}

We have conducted a first overview of cosmic ray interactions with the solar surface via a self-consistent MHD model, Bifrost.
The particular model chosen describes the dynamic equilibrium in a coronal-hole region, linking the convective interior to the low corona.
We expect \textit{a priori} that the coronal holes, if they contain most of the open flux, would ``attract'' more cosmic-ray flux than would closed-field regions such as streamers.
From the point of view of the St{\o}rmer problem, these regions would have lower cutoff rigidities than weak closed-field regions in the photosphere simply because of particle dynamics. 
From Bifrost's point of view, the relative evacuation of strong-field regions may simply represent the squeezing-out of gas by the natural tendency to restore the hydrostatic equilibrium distorted by convectively driven flows.

This paper has dealt only with the quiet Sun, specifically with a Bifrost model suitable for a coronal-hole region.
This may be appropriate for some aspects of SEPs, as well as GCRs, given the  likelihood that global coronal shock waves accelerate these particles.
If this acceleration takes place on open fields, the SEPs that can return to the Sun and interact there may favor coronal holes (see \cite{2018ApJ...867..122J} and \cite{2018IAUS..335...49H}).
Of course the dominant $\gamma$-ray emission from solar flares also has a clear association with closed magnetic fields
\citep{2003ApJ...595L..77H} associated with large sunspots.
The ``sustained'' emission of high-energy $\gamma$-rays \citep{2018ApJ...869..182S} suggests emissions from still a third geometry (neither the flare itself, nor any shock acceleration distant from the Sun), and so the model considerations could be different from the simple case we have studied here.

\section{Acknowledgments}
HSH thanks the Institute of Theoretical Astrophysics, Oslo, for hospitality, and also thanks Institute members Lyndsay Fletcher and Boris Gudiksen for helpful comments.

\bibliography{zweibel-haber}
\bibliographystyle{mnras}

\end{document}